\documentclass[twocolumn,english,aps,pra,showpacs]{revtex4}
\usepackage[T1]{fontenc}
\usepackage[latin9]{inputenc}
\usepackage{amsmath}
\usepackage{graphicx}
\usepackage{amssymb}
\usepackage{esint}

\makeatletter
\@ifundefined{textcolor}{}
{%
 \definecolor{BLACK}{gray}{0}
 \definecolor{WHITE}{gray}{1}
 \definecolor{RED}{rgb}{1,0,0}
 \definecolor{GREEN}{rgb}{0,1,0}
 \definecolor{BLUE}{rgb}{0,0,1}
 \definecolor{CYAN}{cmyk}{1,0,0,0}
 \definecolor{MAGENTA}{cmyk}{0,1,0,0}
 \definecolor{YELLOW}{cmyk}{0,0,1,0}
 }

\makeatother

\makeatother

\usepackage{babel}

\begin{document}

\title{Fulde-Ferrell superfluidity in ultracold Fermi gases with Rashba
spin-orbit coupling}

\author{Hui Hu$^{1}$ and Xia-Ji Liu$^{1}$}

\email{xiajiliu@swin.edu.au}

\affiliation{$^{1}$Centre for Atom Optics and Ultrafast Spectroscopy, Swinburne
University of Technology, Melbourne 3122, Australia}

\date{\today}
\begin{abstract}
We theoretically investigate the inhomogeneous Fulde-Ferrell (FF)
superfluidity in a three dimensional atomic Fermi gas with Rashba
spin-orbit coupling near a broad Feshbach resonance. We show that
within mean-field theory the FF superfluid state is always more favorable
than the standard Bardeen-Cooper-Schrieffer (BCS) superfluid state
when an in-plane Zeeman field is applied. We present a qualitative
finite-temperature phase diagram near resonance and argue that the
predicted FF superfluid is observable with experimentally attainable
temperatures (i.e., $T\sim0.2T_{F}$, where $T_{F}$ is the characteristic
Fermi degenerate temperature). 
\end{abstract}

\pacs{05.30.Fk, 03.75.Hh, 03.75.Ss, 67.85.-d}

\maketitle

\section{Introduction}

The Fulde-Ferrell (FF) superfluid is a fascinating state proposed
to understand the fermionic superfluidity with unequal populations
in the two spin states \cite{Fulde1964}. Unlike the standard Bardeen-Cooper-Schrieffer
(BCS) superfluid, where fermions of opposite spin and momentum form
Cooper pairs and condense into a microscopic state at rest, the FF
superfluid is characterized by Cooper pairs carrying a single-valued
center-of-mass momentum and thus by an inhomogeneous condensate state.
More complicated inhomogeneous condensate states are also possible
with the inclusion of more center-of-mass momenta for their spatial
structure in real space, as suggested by Larkin and Ovchinnikov (LO)
\cite{Larkin1964}. These forms of inhomogeneous superfluidity have
now been collectively referred to as FFLO superfluids. They are anticipated
to have manifestations in a number of important physical settings,
ranging from solid-state superconductors to the nuclear matter at
the heart of neutron stars \cite{Casalbuoni2004,Radzihovsky2010,Gubbels2013}.
However, despite tremendous theoretical and experimental efforts over
the past fifty years, conclusive experimental evidence of their existence
remains elusive \cite{Casalbuoni2004,Radzihovsky2010,Gubbels2013,Radovan2003,Uji2006,Kenzelmann2008,Sheehy2006,He2006,Hu2006,Orso2007,Hu2007,Guan2007,Liu2007,Liu2008,Liao2010}.
In this work, we show that the long-sought FF superfluid might be
observable in a three dimensional (3D) ultracold atomic Fermi gas
with Rashba spin-orbit coupling and in-plane Zeeman field.

\begin{figure}
\begin{centering}
\includegraphics[clip,width=0.48\textwidth]{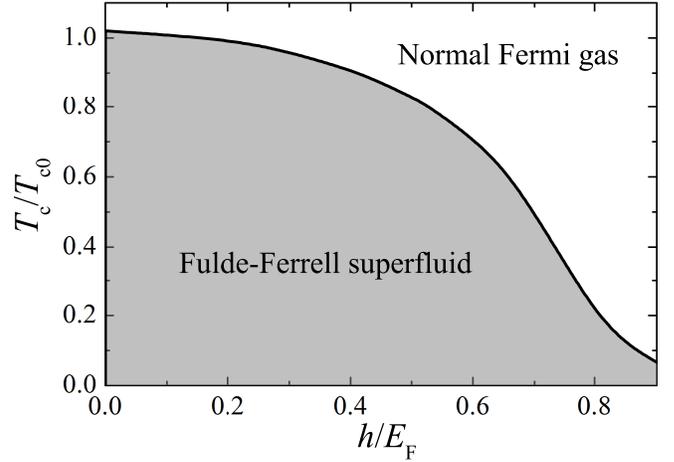} 
\par\end{centering}

\caption{(color online) Finite temperature phase diagram of a 3D Rashba spin-orbit
coupled atomic Fermi gas at a broad Feshbach resonance ($1/k_{F}a_{s}=0$)
and at the spin-orbit coupling strength $\lambda k_{F}/E_{F}=1$.
Here, $k_{F}=(3\pi^{2}n_{F})^{1/3}$ and $E_{F}=\hbar^{2}k_{F}^{2}/(2m)$
are the Fermi wavevector and Fermi energy, respectively, expressed
in terms of the gas density $n_{F}$. The critical temperature, in
units of the mean-field critical temperature without spin-orbit coupling
$T_{c0}\simeq0.496T_{F}=0.496E_{F}/k_{B}$, decreases monotonically
with increasing the in-plane Zeeman field $h$.}

\label{fig1} 
\end{figure}

The idea that inhomogeneous superfluidity is enhanced by Rashba spin-orbit
coupling was first suggested by Barzykin and Gor'kov in the study
of surface superfluidity in materials such as WO$_{3}$:Na \cite{Barzykin2002,Dimitrova2007}
and was later generalized to a 3D Fermi system by Agterberg and Kaur
\cite{Agterberg2007}. It was recently revisited by Zheng and co-workers
in the context of ultracold atomic Fermi gases \cite{Zheng2012a,Zheng2012b},
which have the unique experimental advantage of unprecedented controllability
in interactions, spin-populations and purity \cite{Bloch2008}. In
contrast to solid-state superconductors, ultracold atomic Fermi gases
are generally prepared in the strongly interacting regime \cite{Gubbels2013},
the so-called crossover regime from a Bose-Einstein condensate (BEC)
to a BCS superfluid, in order to have an experimentally attainable
superfluid transition temperature. As a result, the theoretical investigation
of inhomogeneous superfluidity in such ultracold matter has to rely
on heavy numerical calculations within mean-field theory. In the previous
work by Zheng and co-workers \cite{Zheng2012a}, the FF superfluidity
has been addressed at zero temperature along the BEC-BCS crossover.
Here, we present a qualitative mean-field phase diagram at finite
temperatures and show that the FF superfluid state might be within
reach at the typical experimental temperature $T\sim0.2T_{F}\simeq0.4T_{c0}$,
where $T_{F}$ is the Fermi degenerate temperature and $T_{c0}\simeq0.496T_{F}$
is the mean-field critical temperature of a resonantly interacting
(unitary) Fermi gas without spin-orbit coupling. We note that the
mean-field theory gives only a qualitative estimate of the superfluid
transition temperature. For example, for a unitary gas, the mean-field
prediction of $T_{c0}\simeq0.496T_{F}$ is significantly larger than
an experimental measurement $T_{c0}\simeq0.167(13)T_{F}$ \cite{Ku2012}.

Our main result is summarized in Fig. \ref{fig1}, where the phase
transition temperature to a normal state is shown as a function of
the in-plane Zeeman field, for a strongly interacting Rashba spin-orbit
coupled Fermi gas at a broad Feshbach resonance. We find that the
FF superfluid is always the true ground state at a finite Zeeman field.

\begin{figure}
\begin{centering}
\includegraphics[clip,width=0.4\textwidth]{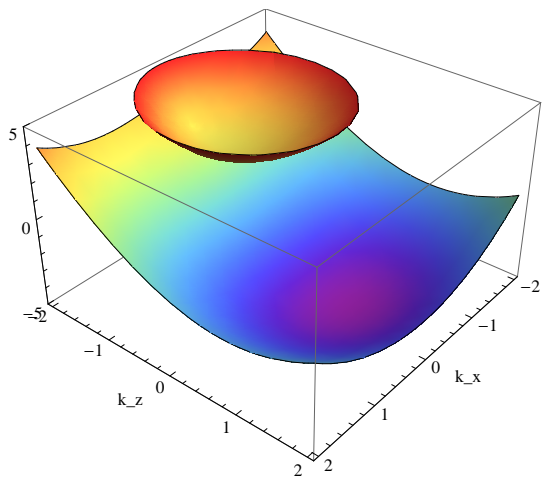} 
\par\end{centering}

\caption{(color online) Schematic of the Fermi surfaces of a Rashba spin-orbit
coupled with in-plane Zeeman field along the \textit{z}-axis.}

\label{fig2} 
\end{figure}

The significantly enlarged parameter space for a FF superfluid can
be qualitatively understood from the change of the two Fermi surfaces
due to spin-orbit coupling, as discussed by Barzykin and Gor'kov \cite{Barzykin2002}
and by Agterberg and Kaur \cite{Agterberg2007}. Here we explain very
briefly the physical picture following their ideas. We consider the
spin-oribt coupling $\lambda(\sigma_{x}\hat{k}_{x}+\sigma_{z}\hat{k}_{z})$
together with an in-plane Zeeman field along the $z$-direction $h\sigma_{z}$,
for which the single-particle energy spectrum takes the form, \begin{equation}
E_{\mathbf{k}\pm}=\frac{\hbar^{2}k^{2}}{2m}\pm\sqrt{\lambda^{2}k_{x}^{2}+(\lambda k_{z}+h)^{2}},\end{equation}
 where {}``$\pm$'' accounts for two helicity branches. The atom
in each branch stays at a \emph{mixed} spin state.

At relatively large spin-orbit coupling (i.e., $\lambda k_{F}\gg h$),
we may approximate \begin{equation}
E_{\mathbf{k}\pm}\simeq\frac{\hbar^{2}\left(k_{x}^{2}+k_{y}^{2}\right)}{2m}+\frac{\hbar^{2}}{2m}\left(k_{z}\pm\frac{q}{2}\right)^{2}\pm\lambda\sqrt{k_{x}^{2}+k_{z}^{2}},\end{equation}
 where \begin{equation}
q=\frac{2mh}{\hbar^{2}\sqrt{k_{x}^{2}+k_{z}^{2}},}.\label{eq:qFF}\end{equation}
 It is easy to see that the centers of the two Fermi surfaces in different
branch are shifted by $(q/2)\mathbf{e}_{z}$ along the $z$-axis in
opposite directions, as shown in Fig. \ref{fig2}, where $\mathbf{e}_{z}$
is the unit vector along the $z$-axis. Thus, if the fermionic pairing
occurs in the lower helicity branch, it would pair two atoms staying
at the single-particle states of $\mathbf{k}+(q/2)\mathbf{e}_{z}$
and $-\mathbf{k}+(q/2)\mathbf{e}_{z}$, respectively, giving rise
to a FF order parameter that has a spatial dependence $\Delta(\mathbf{x})=\Delta e^{iqz}$
\cite{Agterberg2007}. This pairing mechanism holds for arbitrary
small in-plane Zeeman field and explains why the FF superfluid is
always more preferable than the standard BCS superfluid at large spin-orbit
coupling. The direction of the FF momentum is uniquely determined
by the form of spin-orbit coupling and its magnitude is roughly proportional
to the in-plane Zeeman field $h$ if $h$ is small, i.e., see Eq.
(\ref{eq:qFF}).

Let us now consider small spin-orbit coupling. In the absence of spin-orbit
coupling, the formation of an inhomogeneous superfluid is driven by
the population imbalance, which also leads to the distortion of the
Fermi surfaces. In that case, there are many equivalent ways to deform
the surfaces. As a result, the direction of the single FF pairing
momentum is not specified. Thus, for any FF superfluid solution with
a pairing momentum $+\mathbf{q}$, we can always find another degenerate
solution with the pairing momentum $-\mathbf{q}$. This indicates
that a stripe LO phase with an order parameter in the form of $\cos(\mathbf{q}\cdot\mathbf{x})$
will be more favorable, which is simply a superposition of the $+\mathbf{q}$
and $-\mathbf{q}$ plane waves. The investigation of the LO phase
in a 3D Fermi gas without spin-orbit coupling has been carried out
by Burkhardt and Rainer many years ago \cite{Burkhardt1994} and recently
by a number of authors \cite{Yoshida2007,Bulgac2008}. In the presence
of spin-orbit coupling, the two solutions with the $+\mathbf{q}$
and $-\mathbf{q}$ plane waves are no longer degenerate. Numerically,
we find that one solution becomes more favorable and the energy difference
between the two solutions increases rapidly with increasing spin-orbit
coupling. Therefore, with increasing spin-orbit coupling, we anticipate
the LO phase will cease to exist and the large spin-orbit coupling
will uniquely determine a single-valued pairing momentum and lead
to a FF superfluid. The competition between LO and FF phases in a
Rashba spin-orbit coupled 3D Fermi gas was recently investigated by
Agterberg and Kaur \cite{Agterberg2007}. The stripe LO phase was
found to cease to exist at large population imbalance.

It is important to note that the experimentally realized spin-orbit
coupling is not of the pure Rashba type \cite{Wang2012,Cheuk2012}.
Instead, it is an equal weight combination of the Rashba and Dresselhaus
spin-orbit couplings (for a detailed discussion, see for example,
Ref. \cite{Hu2012}). The possibility of observing FF superfluidity
in current experimental settings has been discussed by Shenoy \cite{Shenoy2012},
the present authors \cite{Liu2013}, and also by Wu and co-workers
but in two dimensions \cite{Wu2013}. We anticipate that the Rashba
spin-orbit coupling might be experimentally realized soon \cite{Sau2011,Anderson2013}.
Furthermore, it is also feasible to create a 3D isotropic spin-orbit
coupling \cite{Anderson2012,Anderson2013}. The FF superfluidity with
3D isotropic spin-orbit coupling has been investigated most recently
by Dong and co-workers \cite{Dong2013a} and by Zhou and colleagues
\cite{Zhou2013}.

The reminder of the paper is organized as follows. In Sec. II we introduce
the model Hamiltonian for a Rashba spin-orbit coupled Fermi gas with
an in-plane Zeeman field and describe the mean-field framework. We
present the explicit expression of the mean-field thermodynamic potential,
with the FF pairing momentum $q$ and pairing order parameter $\Delta$
as the variational parameters. In Sec. III, we discuss in detail competing
ground states near a broad Feshbach resonance and show that the FF
superfluid is always the true ground state at finite Zeeman fields
in the superfluid phase. We explore systematically the properties
of this exotic state of matter at finite temperatures. In Sec. IV
we present our conclusions. For the convenience of numerical calculations,
we use a non-standard form of Rashba spin-orbit coupling. In the Appendix
A, we show that it is fully equivalent to the Rashba spin-orbit coupling
commonly used in the literature.

\section{Model Hamiltonian}

We consider a 3D spin-1/2 Fermi gas of $^{6}$Li or $^{40}$K atoms
near broad Feshbach resonances with Rashba-type spin-orbit coupling
$\lambda(\sigma_{x}\hat{k}_{x}+\sigma_{z}\hat{k}_{z})$ and an in-plane
Zeeman field along the $z$-direction $h\sigma_{z}$, a configuration
to be experimentally realized in the future \cite{Sau2011,Anderson2013}.
Here $\sigma_{x}$ and $\sigma_{z}$ are the Pauli matrices, $\hat{k}_{x}\equiv-i\partial_{x}$
and $\hat{k}_{z}\equiv-i\partial_{z}$ are the momentum operators.
In the Appendix A, we discuss in more detail about the expression
of Rashba spin-orbit coupling. The model Hamiltonian of the system
may be described by \begin{equation}
{\cal H}=\int d{\bf x\left[{\cal H}_{0}\left({\bf x}\right)+{\cal {\cal H}_{\textrm{int}}}\left({\bf x}\right)\right]},\end{equation}
 where the single-particle Hamiltonian takes the form \begin{equation}
{\cal H}_{0}=\left[\psi_{\uparrow}^{\dagger},\psi_{\downarrow}^{\dagger}\right]\left[\begin{array}{cc}
\hat{\xi}_{\mathbf{k}}+\lambda\hat{k}_{z}+h & \lambda\hat{k}_{x}\\
\lambda\hat{k}_{x} & \hat{\xi}_{\mathbf{k}}-\lambda\hat{k}_{z}-h\end{array}\right]\left[\begin{array}{c}
\psi_{\uparrow}\\
\psi_{\downarrow}\end{array}\right]\label{eq:spHami}\end{equation}
 and the pairing interaction Hamiltonian is given by \begin{equation}
{\cal H}_{\textrm{int}}=U_{0}\psi_{\uparrow}^{\dagger}\left({\bf x}\right)\psi_{\downarrow}^{\dagger}\left({\bf x}\right)\psi_{\downarrow}\left({\bf x}\right)\psi_{\uparrow}\left({\bf x}\right),\label{eq:intHami}\end{equation}
 describing the contact interaction between the two spin states with
interaction strength $U_{0}$. In the above Hamiltonian, $\psi_{\sigma}^{\dagger}\left({\bf x}\right)$
and $\psi_{\sigma}\left({\bf x}\right)$ are respectively the creation
and annihilation field operators for atoms in the spin-state $\sigma$,
and $\hat{\xi}_{\mathbf{k}}\equiv-\hbar^{2}\nabla^{2}/(2m)-\mu$ is
the single-particle kinetic energy with atomic mass $m$ in reference
to the chemical potential $\mu$. We have denoted the strength of
Rashba spin-orbit coupling and of in-plane Zeeman field by $\lambda$
and $h$, respectively. The use of the contact interatomic interaction
generally leads to an ultraviolet divergence at large momentum and
high energy. To remove such a divergence, it is useful to express
the interaction strength $U_{0}$ in terms of the \textit{s}-wave
scattering length $a_{s}$, \begin{equation}
\frac{1}{U_{0}}=\frac{m}{4\pi\hbar^{2}a_{s}}-\frac{1}{V}\sum_{{\bf k}}\frac{m}{\hbar^{2}k^{2}},\end{equation}
 where $V$ is the volume of the system. In principle, the scattering
length $a_{s}$ may be tuned precisely to arbitrary value, by sweeping
an external magnetic field across the Feshbach resonance \cite{Bloch2008}.
However, in the proposed schemes for creating Rashba spin-orbit coupling
\cite{Sau2011,Anderson2013}, the magnetic bias field must be fine-tuned
to adjust the energy levels of the hyperfine states. This means that
the scattering length for a particular type of atoms may be restricted
to the weak-coupling regime, in which we know that without spin-orbit
coupling it is difficult to have an experimentally accessible superfluid
transition temperature. Nevertheless, the many-body pairing could
be significantly enhanced by Rashba spin-orbit coupling \cite{Kubasiak2010,Vyasanakere2011,Hu2011,Yu2011,Jiang2011,Vyasanakere2012}.
By properly tuning the Rashba spin-orbit coupling strength, which
in some sense equivalent to tuning the scattering length \cite{Vyasanakere2012},
we do anticipate a sizable superfluid transition temperature.

\subsection{Mean-field Bogoliubov-de Gennes theory}

A solid-state Fermi system with Rashba spin-orbit coupling and in-plane
Zeeman field provides a promising platform to observe the long-sought
FFLO superfluid, as suggested by Barzykin and Gor'kov in their pioneering
work \cite{Barzykin2002}. In the context of ultracold atomic Fermi
gases, this idea was renewed, as motivated by the interesting finding
that the two-body bound state of the model Hamiltonian acquires a
finite center-of-mass momentum along the $z$-axis \cite{Dong2013b}.
This strongly indicates the existence of a FF pairing state with a
single-valued center-of-mass momentum at the many-body level \cite{Shenoy2012}.
Indeed, in a recent zero-temperature calculation by Zheng and co-workers
\cite{Zheng2012a}, the parameter space of the FF superfluid has been
found to be significantly enhanced by the Rashba spin-orbit coupling.
Here we explore the whole mean-field phase diagram at finite temperatures.

Let us assume an order parameter with a single-valued center-of-mass
momentum along the $z$-axis: \begin{equation}
\Delta(\mathbf{x})=-U_{0}\left\langle \psi_{\downarrow}(\mathbf{x})\psi_{\uparrow}(\mathbf{x})\right\rangle =\Delta e^{iqz}.\end{equation}
 The direction of the FF pairing momentum is chosen following the
center-of-mass momentum of the two-particle ground state \cite{Dong2013b}.
It is also consistent with the previous mean-field studies for Rashba
spin-orbit coupled Fermi systems \cite{Barzykin2002,Agterberg2007,Zheng2012a}.
Within mean-field theory, we approximate the interaction Hamiltonian
by, \begin{equation}
{\cal H}_{\textrm{int}}\simeq-\left[\Delta(\mathbf{x})\psi_{\uparrow}^{\dagger}(\mathbf{x})\psi_{\downarrow}^{\dagger}(\mathbf{x})+\textrm{H.c.}\right]-\frac{\Delta^{2}}{U_{0}}.\end{equation}
 For a Fermi superfluid, it is convenient to use the following Nambu
spinor representation for the field operators, \begin{equation}
\Phi(\mathbf{x})\equiv\left[\psi_{\uparrow}\left(\mathbf{x}\right),\psi_{\downarrow}\left(\mathbf{x}\right),\psi_{\uparrow}^{\dagger}\left(\mathbf{x}\right),\psi_{\downarrow}^{\dagger}\left(\mathbf{x}\right)\right]^{T},\end{equation}
 where the first two and last two field operators in $\Phi(\mathbf{x})$
could be interpreted as the annihilation operators for particles and
holes, respectively. The total Hamiltonian can then be written in
a compact form, \begin{equation}
\mathcal{H}=\frac{1}{2}\int d{\bf x}\Phi^{\dagger}(\mathbf{x})\mathcal{H}_{BdG}\Phi(\mathbf{x})-V\frac{\Delta^{2}}{U_{0}}+\sum_{\mathbf{k}}\hat{\xi}_{\mathbf{k}},\label{eq:mf_totHamil}\end{equation}
 where the factor of $1/2$ in the first term arises from the double
use of particle and hole operators in the Nambu spinor $\Phi(\mathbf{x})$.
Accordingly, a zero-point energy $\sum_{\mathbf{k}}\hat{\xi}_{\mathbf{k}}$
appears in the last term, which is formally divergent. The Bogoliubov
Hamiltonian $\mathcal{H}_{BdG}$ takes the form,\begin{widetext}
\begin{equation}
\mathcal{H}_{BdG}\equiv\left[\begin{array}{cccc}
\hat{\xi}_{\mathbf{k}}+\lambda\hat{k}_{z}+h & \lambda\hat{k}_{x} & 0 & -\Delta\left(\mathbf{x}\right)\\
\lambda\hat{k}_{x} & \hat{\xi}_{\mathbf{k}}-\lambda\hat{k}_{z}-h & \Delta\left(\mathbf{x}\right) & 0\\
0 & \Delta^{*}\left(\mathbf{x}\right) & -\hat{\xi}_{\mathbf{k}}+\lambda\hat{k}_{z}-h & \lambda\hat{k}_{x}\\
-\Delta^{*}\left(\mathbf{x}\right) & 0 & \lambda\hat{k}_{x} & -\hat{\xi}_{\mathbf{k}}-\lambda\hat{k}_{z}+h\end{array}\right].\label{eq:BdGHami}\end{equation}

In free space, where the momentum is a good quantum number, it is
straightforward to diagonalize the Bogoliubov Hamiltonian \begin{equation}
\mathcal{H}_{BdG}\Phi_{\mathbf{k\eta}}(\mathbf{x})=E_{\mathbf{k\eta}}\Phi_{\mathbf{k\eta}}(\mathbf{x}),\end{equation}
 by using the plane-wave quasiparticle wave-function \begin{equation}
\Phi_{\mathbf{k\eta}}(\mathbf{x})=\frac{1}{\sqrt{V}}e^{i\mathbf{k\cdot}\mathbf{x}}[u_{\mathbf{k\eta\uparrow}}e^{+iqz/2},u_{\mathbf{k\eta\downarrow}}e^{+iqz/2},v_{\mathbf{k\eta\uparrow}}e^{-iqz/2},v_{\mathbf{k}\eta\downarrow}e^{-iqz/2}]^{T}\end{equation}
 and quasiparticle energy $E_{\mathbf{k}\eta}$. The Bogoliubov Hamiltonian
now becomes a 4 by 4 matrix and the four eigenvalues and eigenstates
have been specified using the index $\eta$ ($\eta=1,2,3,4$).

The mean-field thermodynamic potential $\Omega$ at a temperature
$T$ is then given by \begin{eqnarray}
\frac{\Omega}{V} & = & \frac{1}{2V}\left[\sum_{\mathbf{k}}\left(\xi_{\mathbf{k}+\mathbf{q}/2}+\xi_{\mathbf{k}-\mathbf{q}/2}\right)-\sum_{\mathbf{k\eta}}E_{\mathbf{k}\eta}\right]-\frac{k_{B}T}{V}\sum_{\mathbf{k\eta}}\ln\left(1+e^{-E_{\mathbf{k}\eta}/k_{B}T}\right)-\frac{\Delta^{2}}{U_{0}},\label{eq:mf_Omega}\end{eqnarray}
 \end{widetext}where the zero-point energy $\sum_{\mathbf{k\eta}}E_{\mathbf{k}\eta}$
in the first term (i.e., the square bracket) is again due to the double
use of particle and hole operators, and the zero-point energy $\sum_{\mathbf{k}}\hat{\xi}_{\mathbf{k}}$
has been rewritten as $\sum_{\mathbf{k}}(\xi_{\mathbf{k}+\mathbf{q}/2}+\xi_{\mathbf{k}-\mathbf{q}/2})/2$
to cancel the divergence in $\sum_{\mathbf{k\eta}}E_{\mathbf{k}\eta}$.
The second term in the above thermodynamic potential accounts for
the thermal excitations of Bogoliubov quasiparticles, which do not
interact with each other in the mean-field approximation. It should
be noted that, the summation over the quasiparticle energy in $\sum_{\mathbf{k\eta}}$
must be restricted to $E_{\mathbf{k}\eta}\geq0$, because of an inherent
particle-hole symmetry in the Nambu spinor representation. For instance,
it is straightforward to check that for any particle state $[u_{\mathbf{k}\uparrow},u_{\mathbf{k}\downarrow},v_{\mathbf{k}\uparrow},v_{\mathbf{k}\downarrow}]$$^{T}$
with energy $E_{\mathbf{k}}\geq0$, there is a one-to-one corresponding
hole state $[v_{-\mathbf{k}\uparrow}^{*},v_{-\mathbf{k}\downarrow}^{*},u_{-\mathbf{k}\uparrow}^{*},u_{-\mathbf{k}\downarrow}^{*}]$$^{T}$
with energy $-E_{\mathbf{k}}$. These two states correspond to the
same physical solution.

It is easy to show that, in the absence of spin-orbit coupling ($\lambda=0$),
we may explicitly write down the expression for the quasiparticle
energy $E_{\mathbf{k}\eta}$. Eq. (\ref{eq:mf_Omega}) then recovers
the thermodynamic potential of a spin-imbalanced 3D Fermi gas \cite{Hu2006}.

\subsection{Numerical solutions}

For a given set of parameters (i.e., the temperature $T$, the interaction
parameter $1/k_{F}a_{s}$ etc.), we solve the order parameter $\Delta$
and the FF pairing momentum $q$ by using the self-consistent stationary
conditions: $\partial\Omega/\partial\Delta=0$ and $\partial\Omega/\partial q=0$,
together with the number equation $N=-\partial\Omega/\partial\mu$
for the chemical potential $\mu$. Different saddle-point solutions
of these coupled equations give competing ground states. At finite
temperatures, the true ground state is the one that has the lowest
free energy $F=\Omega+\mu N$.

In numerical calculations, we take the Fermi wave-vector $k_{F}=(3\pi^{2}n_{F})^{1/3}$
and the Fermi energy $E_{F}=\hbar^{2}k_{F}^{2}/(2m)$ as the units
for wave-vector and energy, respectively, in order to make the equations
dimensionless. Here, $n_{F}=N/V$ is the gas density. We focus on
the unitary limit with a divergent scattering length $1/(k_{F}a_{s})=0$.
Throughout the paper, we shall use a Rashba spin-orbit coupling strength
$\lambda k_{F}/E_{F}=1$. We will consider the superfluid transition
temperature as a function of the in-plane Zeeman field, as well as
the critical Zeeman field across the Feshbach resonance at a given
temperature.

\section{Results and discussions}

\begin{figure}
\begin{centering}
\includegraphics[clip,width=0.48\textwidth]{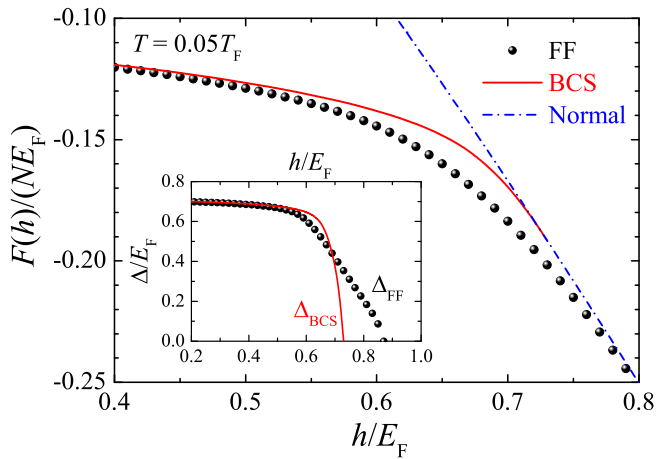} 
\par\end{centering}

\caption{(color online) The free energy of three competing states, including
(i) a normal Fermi gas with $\Delta=0$; (ii) a fully paired BCS superfluid
with $\Delta\neq0$ and $q=0$; and (iii) a finite-momentum paired
FF superfluid with $\Delta\neq0$ and $q\neq0$, as a function of
the in-plane Zeeman field at resonance and at $T=0.05T_{F}$. The
Rashba spin-orbit coupling strength is $\lambda k_{F}/E_{F}=1$. The
inset shows the Zeeman field dependence of the BCS (red solid line)
and FF order parameters (black solid circles).}

\label{fig3} 
\end{figure}

For any set of parameters, in general there are three competing states:
the normal gas ($\Delta=0$), BCS superfluid ($\Delta\neq0$ and $q=0$),
and FF superfluid ($\Delta\neq0$ and $q\neq0$). These competing
states all satisfy the stability condition $\partial^{2}\Omega/\partial\Delta^{2}\geq0$
and therefore are stable against phase separation in real space. In
Fig. \ref{fig3}, we show the free energy of these states in the unitary
limit at $T=0.05T_{F}$. It is readily seen that the FF superfluid
is always more favorable in energy than the standard BCS superfluid
at a finite in-plane Zeeman field, when the Rashba spin-orbit coupling
is present. This observation holds for any interaction parameters
and temperatures.

At small in-plane Zeeman fields (i.e., $h<0.5E_{F}$), the FF superfluid
and BCS superfluid are very similar, with essentially the same pairing
gap, as shown in the inset of Fig. \ref{fig3}. The difference in
the free energy between these two superfluids is indeed negligible.
However, strictly speaking, we always find $\partial\Omega/\partial q<0$
for the BCS superfluid, indicating that a FF superfluid with a finite
center-of-mass momentum $q$ is preferable, although the magnitude
of $q$ might be very small. With increasing the Zeeman field, the
BCS superfluid ceases to exist. The FF superfluid will also disappear,
but at a bit larger critical Zeeman field. This is the so-called Chandrasekhar-Clogston
(CC) limit \cite{Chandrasekhar1962,Clogston1962}, above which a superfluid
breaks down. The FF pairing gap vanishes much more smoothly than the
BCS pairing gap, as can be seen from the inset of Fig. \ref{fig3}.
The phase transition from the FF superfluid to the normal state is
continuous.

By collecting the in-plane CC Zeeman field for the FF superfluid state
at different finite temperatures, we obtain a finite-temperature phase
diagram in the strongly interacting unitary limit, as shown in Fig.
\ref{fig1}. For a 2D weakly interacting Fermi gas with Rashba spin-orbit
coupling, such a phase diagram has been determined by Barzykin and
Gor'kov \cite{Barzykin2002}. At small Zeeman fields $h\sim0$, we
find that \begin{equation}
\left[T_{c}\left(h=0\right)-T_{c}\right]/T_{c}\left(h=0\right)\propto h^{2},\end{equation}
 in agreement with the analytic expression in Ref. \cite{Barzykin2002}.
At the low-temperature regime, the transition temperature $T_{c}(h)$
is roughly given by \begin{equation}
T_{c}\left(h\right)\propto\left[h_{c}\left(T=0\right)-h\right]{}^{\alpha},\end{equation}
 where $\alpha\simeq2.5$ and $h_{c}(T=0)\simeq1.05E_{F}$ is the
CC Zeeman field at zero temperature. Our result of $\alpha\simeq2.5$
seems to be consistent with the analytic prediction of $\alpha=3$
for a 2D Rashba spin-orbit coupled Fermi gas \cite{Barzykin2002}.

In the absence of spin-orbit coupling, the CC Zeeman field at zero
temperature is given by $h_{c}\simeq1/\sqrt{2}\Delta\simeq0.707\Delta$
(or $h_{c}\simeq0.754\Delta$ with the inclusion of the possibility
of a FF superfluid) in the weakly-interacting regime. This is obtain
by relating the free energy of a superfluid $F_{S}$ and of a normal
gas $F_{N}$ through \cite{Clogston1962}, \begin{equation}
F_{N}-F_{S}=\frac{1}{2}\chi h^{2},\label{eq:ClogstonEq}\end{equation}
 where $\chi$ is the spin susceptibility of a normal gas. In the
unitary limit, where the pairing gap $\Delta$ is comparable with
the Fermi energy $E_{F}$, the similar argument within mean-field
theory predicts $h_{c}\simeq0.693E_{F}$ \cite{Sheehy2006}. In the
presence of spin-orbit coupling, our result of $h_{c}\simeq1.05E_{F}$
seems to be significantly larger. This enhancement of the CC Zeeman
field may be understood from the decrease of susceptibility due to
Rashba spin-orbit coupling, i.e., it is reduced by a factor of $2$
in the weak coupling limit \cite{Gorkov2001}. As shown by Clogston,
if the susceptibility is reduced by a fraction $\alpha$, the CC limit
should be divided by $1/\sqrt{\alpha}$ \cite{Clogston1962}. Thus,
by setting $\alpha=1/2$, the argument by Chandrasekhar and Clogston
leads to $h_{c}\simeq0.693\sqrt{2}E_{F}\simeq0.98E_{F}$, in a good
agreement with our numerical calculation.

It is important to note that our phase diagram in Fig. \ref{fig1}
seems to be qualitatively different from the one obtained in the previous
study by Zheng and collaborators, which predicts a BCS superfluid
at small in-plane Zeeman field \cite{Zheng2012a}. In addition, the
CC Zeeman field at zero temperature $h_{c}(T=0)$ is shown in their
Fig. 1(d) to be about $0.55E_{F}$ \cite{Zheng2012a}, much smaller
than what we have obtained numerically, i.e., $h_{c}(T=0)\sim1.05E_{F}$.
The qualitatively different phase diagram is simply due to the different
interpretation of the FF state with small pairing momentum. For example,
in the calculations by Zheng and collaborators, the FF state with
pairing momentum $q<10^{-3}k_{F}$ has been regarded as the BCS superfluid
\cite{ZhengPrivateCommunication}. On the other hand, the different
CC Zeeman field is probably due to the different accuracy of numerical
calculations \cite{ZhengPrivateCommunication}.

\begin{figure}
\begin{centering}
\includegraphics[clip,width=0.48\textwidth]{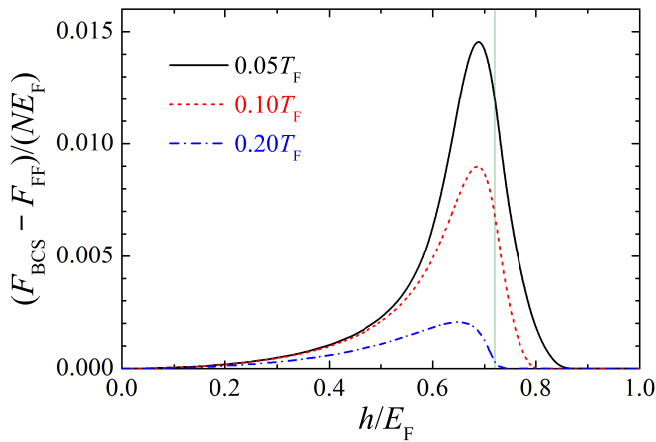} 
\par\end{centering}

\caption{(color online) The free energy difference between the FF superfluid
state and the possible BCS superfluid state at resonance and at three
different temperatures, $T=0.05T_{F}$ (black solid line), $0.10T_{F}$
(red dashed line), and $0.20T_{F}$ (blue dot-dashed line). The Rashba
spin-orbit coupling strength is $\lambda k_{F}/E_{F}=1$. The BCS
state ceases to exist above a Zeeman field $h_{BCS,N}\simeq0.72E_{F}$,
which depends very weakly on temperature and is indicated by the thin
vertical line in the figure. At $h>h_{BCS,N}$, we replace $F_{BCS}$
by $F_{N}$ in the calculation of the free energy difference.}

\label{fig4} 
\end{figure}

\subsection{Temperature dependence of the FF superfluid}

We now explore in greater detail the finite-temperature properties
of the FF superfluid. Fig. \ref{fig4} reports the difference in free
energy between the FF superfluid state and the BCS superfluid state
(the normal state) at three typical temperatures \cite{Ku2012}. The
energy difference per particle is sizable at low temperatures, suggesting
that the FF superfluid is very robust with respect to other competing
ground states. The difference becomes considerably smaller with increasing
temperature. Nevertheless, it is still visible (i.e., at about $0.002E_{F}$)
at the typical experimental temperature $T=0.2T_{F}$. Thus, the thermodynamic
stability of the FF superfluid could be guaranteed. More accurately,
one may use the so-called Thouless criterion generalized to allow
the determination of the FF pairing instability at finite temperatures
\cite{Liu2006EPL}. Our preliminary result, which partially takes
into account the strong pair fluctuations (not shown in the figure),
indicates that indeed with decreasing temperature the pairing instability
of a normal Fermi gas always occurs at a finite center-of-mass momentum.
This is consistent with the mean-field prediction shown in Fig. \ref{fig4}.

\begin{figure}
\begin{centering}
\includegraphics[clip,width=0.48\textwidth]{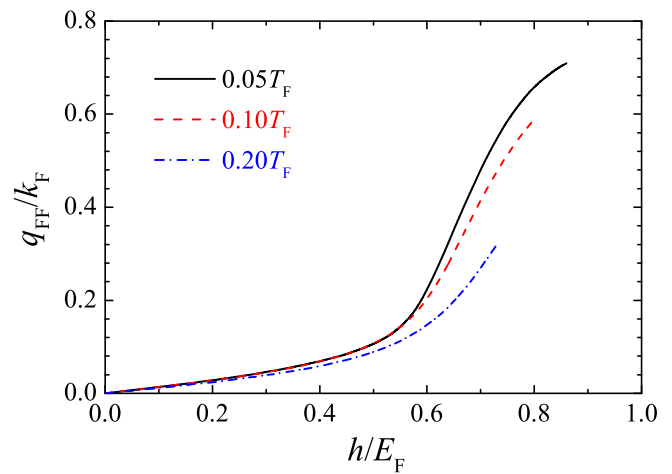} 
\par\end{centering}

\caption{(color online) The pairing momentum of the FF superfluid at resonance
and at three different temperatures, $T=0.05T_{F}$ (black solid line),
$0.10T_{F}$ (red dashed line), and $0.20T_{F}$ (blue dot-dashed
line). The Rashba spin-orbit coupling strength is $\lambda k_{F}/E_{F}=1$.}

\label{fig5} 
\end{figure}

Fig. \ref{fig5} shows the FF pairing momentum as a function of the
in-plane Zeeman field at different temperatures. At low Zeeman fields,
the pairing momentum is essentially independent on the temperature.
At high fields about $0.6E_{F}$, however, the pairing momentum decreases
quickly with increasing temperature. We note that, near zero temperature,
the maximum pairing momentum is comparable to the Fermi wave-vector
$k_{F}$.

\begin{figure}
\begin{centering}
\includegraphics[clip,width=0.4\textwidth]{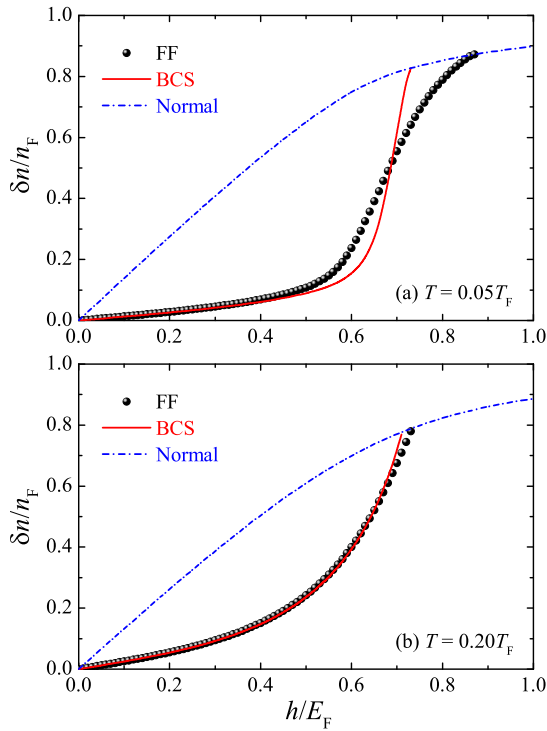} 
\par\end{centering}

\caption{(color online) The number difference between the two spin states at
$T=0.05T_{F}$ (upper panel) and $T=0.20T_{F}$ (lower panel) as a
function of the in-plane Zeeman field at resonance and at the spin-orbit
coupling strength $\lambda k_{F}/E_{F}=1$. In each panel, the number
difference of the three competing states is shown by solid circles
(FF), red line (BCS), and blue dot-dashed line (normal Fermi gas).}

\label{fig6} 
\end{figure}

Fig. \ref{fig6} presents the population imbalance between the two
spin states, calculated by using $V\delta n=-\partial\Omega/\partial h$
and normalized by the gas density $n_{F}\equiv k_{F}^{3}/(3\pi^{2})$.
The BCS superfluid state has low capacity to accommodate the population
imbalance. Thus, it ceases to exist at a threshold Zeeman field. The
deformation of the Fermi surfaces in the FF superfluid state is able
to allow more population imbalances. At low temperatures (upper panel,
$T=0.05T_{F}$), we observe that the population imbalance of the BCS
superfluid rises quickly near its threshold Zeeman field, compared
with that of the FF superfluid near the CC limit. This distinct behavior
might be used to identify the FF superfluid. At a relatively large
temperature (low panel, $T=0.2T_{F}$), the difference in the population
imbalance between the two superfluids is smeared out by temperature.

\begin{figure}
\begin{centering}
\includegraphics[clip,width=0.48\textwidth]{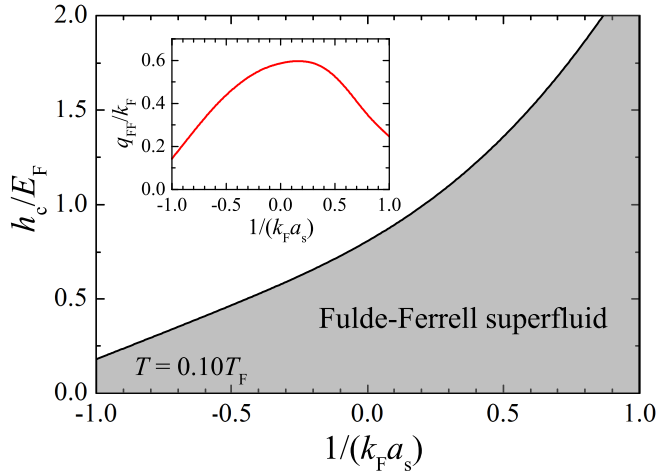} 
\par\end{centering}

\caption{(color online) Phase diagram along the BEC-BCS crossover at $T=0.10T_{F}$
and $\lambda k_{F}/E_{F}=1$. The main figure reports the CC Zeeman
field at which the FF superfluid phase turns into the normal state.
The inset shows the FF pairing momentum at the CC Zeeman field. This
figure might be contrasted with the zero temperature result in Ref.
\cite{Zheng2012a}, see for example, Fig. 1(c) in Ref. \cite{Zheng2012a}.}

\label{fig7} 
\end{figure}

\subsection{FF superfluid across the BEC-BCS crossover}

We have so far focused on the resonance limit with $1/(k_{F}a_{s})=0$.
In Fig. \ref{fig7}, we show the CC Zeeman field for the FF superfluid
state across the BEC-BCS crossover at the temperature $T=0.1T_{F}$.
It increases monotonically when the Fermi cloud crosses from the BCS
limit to the BEC limit. In the inset, we present the FF pairing momentum
at the CC Zeeman field. There is a maximum in the pairing momentum
at about the Feshbach resonance, due to the competition between the
many-body and two-body effects. The initial increase of the FF pairing
momentum on the BCS side arises from the many-body effect. Towards
the BEC limit, however, the two-body effect becomes dominant and the
FF pairing momentum decreases gradually to the center-of-mass momentum
of the two-particle bound state. In the previous study by Dong and
collaborators, the many-body and two-body predictions for the FF pairing
momentum have been compared with each other for a 3D Fermi gas with
3D isotropic spin-orbit coupling \cite{Dong2013a}. Our result in
the inset is consistent with their predictions.

\section{Conclusions}

In summary, we have investigated the finite-temperature phase diagram
of a three dimensional atomic Fermi gas with Rashba spin-orbit coupling
and in-plane Zeeman field near a broad Feshbach resonance. We have
shown that its superfluid state is always an inhomogeneous Fulde-Ferrell
superfluid, if the spin-orbit coupling is sufficiently large.

Our work complements the previous studies with spin-orbit coupled
Fermi systems in the solid-state, which are described by essentially
the same model Hamiltonian in the weakly interacting regime \cite{Barzykin2002,Dimitrova2007,Agterberg2007}.
Our mean-field treatment gives a qualitative picture of the Fulde-Ferrell
superfluidity in the strongly interacting resonance limit, which is
of great interest and of experimental relevance.

Our result extends the previous zero-temperature investigation by
Zheng and collaborators \cite{Zheng2012a} to finite temperatures.
In particular, we have clarified that the BCS superfluid found by
these authors at small in-plane Zeeman fields is better understood
as the Fulde-Ferrell superfluid with small pairing momentum. Our finite-temperature
calculations indicate that the Fulde-Ferrell superfluidity could be
observable at about $T\sim0.2T_{F}$, a temperature that is already
reached in current cold-atom experiments when the spin-orbit coupling
is absent \cite{Ku2012}.

Our investigation is based on the mean-field theory, which is known
to provide qualitative picture of the BEC-BCS crossover. The strong
pair fluctuations in a strongly interacting Fulde-Ferrell superfluid
might be taken into account by using many-body $T$-matrix theories
\cite{Liu2006EPL,Hu2006EPL,Hu2008TmatrixPRA}. This is to be addressed
in the future work. Our preliminary analysis indicates that the superfluid
transition temperature predicted in Fig. \ref{fig1} will decrease
by a factor of about $2$ at the typical in-plane Zeeman field $h=0.6E_{F}$. 
\begin{acknowledgments}
We are grateful to Han Pu and Lin Dong for stimulating discussions,
and Zhen Zheng and Xubo Zou for helpful communications. H.H. thanks
the hospitality of Physics Department, East China Normal University,
where this manuscript was completed. This research was supported by
the ARC Discovery Projects (DP0984522 and DP0984637) and the NFRP-China
2011CB921502. 
\end{acknowledgments}
\appendix

\section{Rashba spin-orbit Hamiltonian}

In the literature, the Rashba spin-orbit coupling and in-plane Zeeman
field is commonly written as $\lambda(\sigma_{y}\hat{k}_{x}-\sigma_{x}\hat{k}_{y})-h\sigma_{x}$
\cite{Zheng2012a}. Here we show that our single-particle Hamiltonian
Eq. (\ref{eq:spHami}) uses exactly the same Rashba-type spin-orbit
coupling and in-plane Zeeman field, after some rotations in real space
or spin space.

For this purpose, we first perform the rotation of real space along
the $x$-axis: $x\rightarrow x$, $y\rightarrow-z$, and $z\rightarrow y$.
As the result, the single-particle Hamiltonian Eq. (\ref{eq:spHami})
becomes, \begin{equation}
\left[\psi_{\uparrow}^{\dagger},\psi_{\downarrow}^{\dagger}\right]\left[\begin{array}{cc}
\hat{\xi}_{\mathbf{k}}+\lambda\hat{k}_{y}+h & \lambda\hat{k}_{x}\\
\lambda\hat{k}_{x} & \hat{\xi}_{\mathbf{k}}-\lambda\hat{k}_{y}-h\end{array}\right]\left[\begin{array}{c}
\psi_{\uparrow}\\
\psi_{\downarrow}\end{array}\right].\end{equation}
 In the second step, let us take a unitary transformation, \begin{align}
\psi_{\uparrow}\left(\mathbf{x}\right) & =\frac{1}{\sqrt{2}}\left[\phi_{\uparrow}\left(\mathbf{x}\right)-\phi_{\downarrow}\left(\mathbf{x}\right)\right],\\
\psi_{\downarrow}\left(\mathbf{x}\right) & =\frac{1}{\sqrt{2}}\left(-i\right)\left[\phi_{\uparrow}\left(\mathbf{x}\right)+\phi_{\downarrow}\left(\mathbf{x}\right)\right].\end{align}
 It is straightforward to check that under such a transformation,
the single-particle Hamiltonian changes to,

\begin{equation}
\left[\phi_{\uparrow}^{\dagger},\phi_{\downarrow}^{\dagger}\right]\left[\begin{array}{cc}
\hat{\xi}_{\mathbf{k}} & -i\lambda\hat{k}_{x}-\lambda\hat{k}_{y}-h\\
i\lambda\hat{k}_{x}-\lambda\hat{k}_{y}-h & \hat{\xi}_{\mathbf{k}}\end{array}\right]\left[\begin{array}{c}
\phi_{\uparrow}\\
\phi_{\downarrow}\end{array}\right],\end{equation}
 which precisely has the form of $\lambda(\sigma_{y}\hat{k}_{x}-\sigma_{x}\hat{k}_{y})-h\sigma_{x}$.

The interaction Hamiltonian Eq. (\ref{eq:intHami}) is not apparently
affected by the spatial rotation. Furthermore, as $\psi_{\downarrow}\left(\mathbf{x}\right)\psi_{\uparrow}\left(\mathbf{x}\right)=-i\phi_{\downarrow}\left(\mathbf{x}\right)\phi_{\uparrow}\left(\mathbf{x}\right)$,
it is also not affected by the unitary transformation. For the FF
pairing momentum, we can see that the direction along the $z$-axis
is fully equivalent to the direction of $y$-axis after the two rotations.
Thus, our assumed $z$-axis for the FF pairing momentum is consistent
with the observation by Zheng and collaborators \cite{Zheng2012a}
and by Barzykin and Gor'kov \cite{Barzykin2002}, that the FF pairing
momentum with $\lambda(\sigma_{y}\hat{k}_{x}-\sigma_{x}\hat{k}_{y})-h\sigma_{x}$
is along the $y$-axis.

\end{document}